\documentclass[sort&compress]{elsarticle}
\usepackage{amsmath,mathtools,amssymb}
\usepackage{graphicx}   
\usepackage{epstopdf}
\usepackage{subfigure}
\usepackage{color}
\usepackage{verbatim} 
\usepackage{booktabs}
\usepackage{bm}   
\usepackage{dutchcal}
\usepackage[left=3cm,right=3cm,top=2.5cm,bottom=2.5cm]{geometry}  
\usepackage{lineno,hyperref}
\usepackage{tablists} 
\usepackage{makecell} 
\usepackage{tikz}
\usetikzlibrary{arrows.meta, positioning}
\usepackage{hyperref}
\usepackage{pstricks}
\usepackage[titletoc]{appendix}
\usepackage{pstricks-add}
\usepackage{float}
\usepackage{mathrsfs} 
\usepackage{amsthm}
\hypersetup{
	colorlinks=ture,
	linkcolor=blue,
	filecolor=gray,
	urlcolor=blue,
	citecolor=blue}
\numberwithin{equation}{section}

\allowdisplaybreaks[4]

\begin{document}
	\begin{frontmatter}
		\title{Extended Cartan homotopy formula for higher Chern-Simons-Antoniadis-Savvidy theory} 
	
		\author{Danhua Song}
		\ead{danhua_song@163.com}
		\address{School of Mathematical Sciences, Capital Normal University, Beijing 100048, China}
		\date{}
		
		\begin{abstract} 
			We consider extended Cartan homotopy formula (ECHF) for higher gauge theory. Firstly, we construct an oriented simplex based on 2-connections and present differential and integral forms of the higher ECHF. Then, we study the higher Chern-Simons-Antoniadis-Savvidy (ChSAS) theory and prove that the higher ECHF can reproduce the higher Chern-Weil theorem and give higher triangle equation. We finally conclude from the higher ECHF that a higher transgression form can be written as the difference of two higher ChSAS forms minus an exact form.
			
		\end{abstract}
		
	\end{frontmatter}
	%
	

	\section{Introduction}
	The extended Cartan homotopy formula (ECHF) \cite{Zumino, FIPSPSS, PRRJ, FEP} is a widely used tool in mathematics and physics,  especially when dealing with Chern-Simons (CS) theory and subspace structures of gauge algebras.
	For instance, the first well known particular case of the ECHF is the Chern-Weil theorem, which is an important content in differential geometry and topology. 
	Besides, the iterative use of the ECHF  gives a subspace separation method, which allows one to (1) separate the CS action in bulk and boundary contributions, and (2) split the Lagrangian in appropriate reflection of the subspace structure of the gauge algebra, systematically.  The method is built upon the triangle equation, which is a corollary of the ECHF.
	 It is worth pointing out that, in order to apply the method, one must regard CS forms as particular cases of more general objects known as transgression forms.  This gives new insight into the CS theory and is also of physical interest, particularly in the context of gravity\cite{FIERPS, PRRJ}.  We are thus led to a result: there is a common origin of the Chern-Weil theorem and the triangle equation.
	This paper aims to develop the ECHF  to higher gauge theory  \cite{Baez.2010, FH, Baez2005HigherGT, 	JBUS, Faria_Martins_2011, doi:10.1063/1.4870640}, and extend the above results of the CS theory to the higher CS theory  \cite{SDH5}. We hope offer a new perspective from which to view the higher gauge theory.
	
	The higher gauge theory is a branch of mathematical physics, which generalizes ordinary gauge theory to higher algebraic structures.
	Based on the development of this theory, many papers have been written about the higher CS theory. In Refs. \cite{ESRZ, R.Z, R.Z2, R.Z3, R.Z4, R.Z5}, Soncini and Zucchini formulated a 4-dimensional  semistrict higher gauge theoretic CS theory. Besides, one studied the higher CS theory arising in the AKSZ-formalism based on the $L_\infty$-algebra \cite{HSUSJS, DFUSJS, DFCLRUS, PRCS}.   Moreover, in Ref. \cite{ SDH4}, the authors constructed low-dimensional 2-Chern-Simons and 3-Chern-Simons gauge theories by applying the generalized differential calculus \cite{Robinson1}.  
	Recently, one developed the $(2n+2)$-dimensional higher CS form, namely the higher Chern-Simons-Antoniadis-Savvidy (ChSAS) form \cite{SDH5}, which is just the research object of this paper.

	In Refs. \cite{GS, IAGS, GS1, SKGS}, Antoniadis,  Konitopoulos and Savvidy   constructed metric independent, gauge invariant, and closed form $\Gamma_{2n+p}$, which is similar to the Pontryagin-Chern form in the ordinary CS theory. The principal significance of $\Gamma_{2n+p}$ is that it can be used in both, odd- and even-dimensional spacetimes. 
	Based on their research, Salgado et al. \cite{FIPSPSS, FIP} discussed a particular extended  invariant form 
	\begin{align}\label{ei}
		\Gamma_{2n+3}= \langle F^n \wedge H\rangle_{\mathcal{g}},
	\end{align}
	where $\langle \cdots \rangle_{\mathcal{g}}$ stands for a multilinear symmetric invariant polynomial  $\langle \cdots \rangle_{\mathcal{g}}: \mathcal{g}^{n+1} \longrightarrow \mathbb{R}$ for the Lie algebra $\mathcal{g}$, and $F=dA+A \wedge A$ is the curvature 2-form for the $\mathcal{g}$-valued connection 1-form $A$ and $H=d B + [A,B]$ is the curvature 3-form for the $\mathcal{g}$-valued connection 2-form field $B$. 
	It is straightforward to show that $d \Gamma_{2n+3}=0$ (see Ref. \cite{IAGS}).
	Then, there is a $(2n+2)$-ChSAS form $\mathfrak{C}^{2n+2}_{ChSAS}$ satisfying $ \Gamma_{2n+3}= d\mathfrak{C}^{2n+2}_{ChSAS}$, which  can be obtained as a special case of the generalized Chern-Weil theorem.
	
	\textbf{Generalized Chern-Weil theorem}:
	Let $(A_0, B_0)$ and $(A_1, B_1)$ be two generalized connections, which consist of the connection 1-forms $A_0$, $A_1$ and connection 2-forms $B_0$, $B_1$, and let $(F_0, H_0)$ and $(F_1, H_1)$ be their generalized curvatures with $F_i = d A_i + A_i \wedge A_i$ and $H_i= dB_i + [A_i, B_i]$ for $i=0, 1$. 
	There are two interpolations $A_t= A_0 + t (A_1- A_0)$ and $B_t= B_0 + t (B_1- B_0)$,  $(0 \leq t \leq 1)$, and the corresponding curvatures are given by $F_t = d A_t + A_t \wedge A_t$ and $H_t= dB_t + [A_t, B_t]$. Then, the difference $\Gamma^{(1)}_{2n+3}- \Gamma^{(0)}_{2n+3}$ is an exact form
	\begin{align}\label{GCW}
	\Gamma^{(1)}_{2n+3}- \Gamma^{(0)}_{2n+3}=\langle F_1^n \wedge H_1\rangle_{\mathcal{g}}-
		\langle F_0^n \wedge H_0\rangle_{\mathcal{g}}=d Q^{2n+2}(A_0, B_0; A_1, B_1),
	\end{align}
	where
	\begin{align}\label{AST}
		Q^{2n+2}(A_0, B_0; A_1, B_1)= \int_{0}^{1}dt\Big\{n\langle F^{n-1}_t \wedge (A_1- A_0) \wedge H_t \rangle_{\mathcal{g}} + \langle F^n_t \wedge (B_1- B_0) \rangle_{\mathcal{g}}\Big\}
	\end{align}
	is called an Antoniadis-Savvidy transgression form.
	
For the proofs we refer the reader to Ref. \cite{FIP}.	Following the same procedure followed in the case of the Chern-Simons forms, the $(2n+2)$-ChSAS form can be given explicitly by
	\begin{align}\label{CASA}
		\mathfrak{C}^{2n+2}_{ChSAS}= Q^{2n+2}(0,0; A, B)=\int_{0}^{1}dt \langle n A \wedge F_t^{n-1}\wedge H_t + F^n_t \wedge B \rangle_{\mathcal{g}}
	\end{align}
by setting $A_0=B_0=0$, $A_1=A$ and $B_1=B$ in \eqref{AST}, and the equation \eqref{GCW} becomes
\begin{align}
	\langle F^n \wedge H\rangle_{\mathcal{g}}=d\mathfrak{C}^{2n+2}_{ChSAS}.
\end{align}
Consequently, the $(2n+2)$-ChSAS form is regarded as a particular case of the Antoniadis-Savvidy transgression form, and it is analogous to the usual Chern-Simons form, but in even dimensions.

	Following Ref. \cite{FIPSPSS}, Salgado et al. presented a generalized ECHF to deal with the ChSAS theory.
	Similar to the usual ECHF, the generalized ECHF can reproduce the generalized Chern-Weil theorem as mentioned above. 
	In addition, there is also a subspace separation method that allows the separation of the ChSAS action into bulk and boundary contributions and the splitting of the bulk Lagrangian into pieces that reflect the particular subspace structure of the gauge algebra.
	By the same token, this method is built upon the generalized triangle equation
	\begin{align}
		Q^{2n+2}(A_0, B_0; A_2, B_2)=&Q^{2n+2}(A_0, B_0; A_1, B_1)+Q^{2n+2}(A_1, B_1; A_2, B_2)\nonumber\\
		&+dQ^{2n+1}(A_0, B_0; A_1, B_1; A_2, B_2),
	\end{align}
where the exact form of the boundary contribution can be given by 
	\begin{align}
		Q^{2n+1}(A_0, B_0; A_1, B_1; A_2, B_2)=&\int_{0}^{1}dt\int_{0}^{t}ds\Big\{n(n-1)\langle F^{n-2}_t \wedge (A_2-A_1)\wedge(A_1-A_0)\wedge H_t\rangle_{\mathcal{g}}\nonumber\\
		&+n\langle F^{n-1}_t \wedge A_0 \wedge (B_2-B_1)\rangle_{\mathcal{g}} + n \langle F^{n-1}_t \wedge A_1 \wedge(B_0-B_2)\rangle_{\mathcal{g}} \nonumber\\
		&+ n \langle F^{n-1}_t \wedge A_2 \wedge (B_1 - B_0)\rangle_{\mathcal{g}}
		\Big\}
	\end{align}
	with $A_t=A_0 +t(A_1-A_0)+s(A_2-A_1)$ and $B_t=B_0 +t(B_1-B_0)+s(B_2-B_1)$ by using the generalized ECHF.  Thus, we are led to a result: there is also a common origin of the generalized Chern-Weil theorem and the generalized triangle equation.
	
	 
	To summarize, we observe that the generalized ECHF in  \cite{FIPSPSS} provides a unified treatment of both the decomposition of gauge structures and topological invariants in ChSAS theory. It not only clearly distinguishes bulk and boundary physical contributions through subspace separation but also maintains the mathematical consistency of the generalized Chern-Weil theorem.  However, in the later work \cite{SDH5}, although higher ChSAS forms were successfully derived using crossed module theory, this approach failed to fully capitalize on the unifying advantages of the ECHF framework. Crucially, the development of higher ECHF and its role in higher ChSAS theory remains an unresolved challenge, forming the key motivation for our current work.
		The central objective of this work is to systematically demonstrate that both classical and generalized versions of the ECHF can be rigorously extended to higher gauge theory. Through the complete mathematical construction of higher ECHF, we establish that this theoretical framework preserves all essential structures parallel to the classical case. Specifically, we prove: 
 	\begin{itemize}
 		\item[1)]the validity of the higher Chern-Weil theorem,
 		\item[2)]the derivation of higher triangle equations, 
 		\item[3)]universal expressions for higher transgression forms. 
 	\end{itemize}
 This generalization ensures the full commutativity of the conceptual development diagram presented in Fig. \ref{fig:path}, thereby providing a unified approach that bridges conventional and higher gauge theories while maintaining their fundamental structural relationships.
\begin{figure}[H]
	\centering
	\begin{tikzpicture}[
		node distance=2cm,
		stage/.style={
			rectangle, 
			draw=red!50, 
			fill=red!10, 
			thick, 
			minimum width=2cm, 
			minimum height=1.2cm, 
			text width=1.8cm,     
			align=center
		},
		theory/.style={
			rectangle,
			draw=green!50,
			fill=green!10,
			thick,
			rounded corners=3pt,  
			minimum width=2cm,
			minimum height=1.2cm,
			text width=1.8cm,
			align=center
		},
		arrow/.style={
			-Stealth, 
			semithick,
			shorten >=2pt, 
			shorten <=2pt
		}
		]
		
		\node[stage] (echf) {Ordinary ECHF};
		\node[stage, right=of echf] (gechf) {Generalized ECHF};
		\node[stage, right=of gechf] (hechf) {Higher ECHF};
		
		\node[theory, below=of echf] (cs) {CS Theory}; 
		\node[theory, below=of gechf] (chsa) {ChSAS Theory};
		\node[theory, below=of hechf] (hchsa) {2-ChSAS Theory};
		
		\draw[arrow] (echf) -- (gechf);
		\draw[arrow] (gechf) -- (hechf);
		
		\draw[arrow] (echf) -- (cs);
		\draw[arrow] (gechf) -- (chsa);
		\draw[arrow] (hechf) -- (hchsa);
		
		\draw[arrow] (cs) to[out=350,in=190] (chsa); 
		\draw[arrow] (chsa) to[out=350,in=190] (hchsa);
	\end{tikzpicture}
	\caption{Development path of ECHF framework and corresponding theories} 
	\label{fig:path}
\end{figure}


This work is organized as follows: in Section \ref{2ChSAS}, we briefly review the principal aspects of the higher ChSAS theory. In Section \ref{hECHF}, we construct the higher ECHF and present higher descent equations, which can be seen as a particular case of the former.
Then, we show that the higher ECHF can reproduce the higher Chern-Weil theorem and give rise to the triangle equation, mimicking what happens in the ordinary and  generalized  ECHF.
In Section \ref{hCHF}, we study the higher Cartan homotopy formula and show that a higher  transgression form can be written as the difference of two higher ChSAS forms minus an exact form.
Lastly, we finish in Section \ref{Concluding} with conclusions and some considerations on future possible developments.

\section{Higher ChSAS forms in $(2n+2)$ dimensions}\label{2ChSAS}
In this section, we will give a brief exposition of the higher ChSAS theory, and set up notations and terminologies which are needed in subsequent sections. For more details, we refer the reader to Ref.  \cite{SDH5}.

Consider a principal 2-bundle $E$ over $M$ which is an oriented, compact manifold without boundary, and let the structure Lie 2-group of $E$ be given in terms of the Lie crossed module $(H, G; \bar{\alpha}, \bar{\vartriangleright})$  with corresponding
differential crossed module $(\mathcal{h}, \mathcal{g}; \alpha, \vartriangleright)$ (see  \ref{crossed module} for the definitions). 
A 2-connection on $E$ is a couple $(A, B)$ with $A=\sum \limits_{a}  A^a X_a\in \Omega^1(M, \mathcal{g})$ and $B=\sum \limits_{b}  B^b Y_b\in \Omega^2(M, \mathcal{h})$, where $A^a$ and $B^b$ are scalar differential $1$- and $2$-form, and
$X_a$ and $Y_b$ are the Lie algebra generator bases of $\mathcal{g}$ and $\mathcal{h}$,  respectively. The corresponding 2-curvature forms $\mathcal{F}\in \Omega^2(M, \mathcal{g})$ and $\mathcal{G}\in \Omega^3(M, \mathcal{h})$ are given by
\begin{align}
	\mathcal{F}= dA + \dfrac{1}{2}A \wedge^{[, ]} A -\alpha(B),\ \ \ 
	\mathcal{G}= dB + A \wedge^{\vartriangleright}B,
\end{align}
which automatically satisfy the 2-Bianchi Identities,
\begin{align} 
	&d \mathcal{F} + A \wedge^{[, ]}\mathcal{F} + \alpha(\mathcal{G})=0,\label{2bi}\\
	& d \mathcal{G} + A \wedge^{\vartriangleright}\mathcal{G} - \mathcal{F}\wedge^{\vartriangleright}B=0.\label{2bi1}
\end{align}
Call $(A, B)$ fake-flat, if $\mathcal{F}=0$, and flat, if it is fake-flat and $\mathcal{G}=0$.
Moreover, there is a general 2-gauge transformation for the 2-connection $(A, B)$,
\begin{align}
	&A'=g^{-1}Ag + g^{-1}dg + \alpha(\phi),\label{2gt11}\\
	&B'=g^{-1}\vartriangleright B + d \phi + A'\wedge^{\vartriangleright}\phi - \phi\wedge\phi,\label{2gt22}
\end{align}
with $g \in G$ and $\phi \in \Omega^1(M, \mathcal{h})$. The corresponding curvatures transform as follows
\begin{align}
	\mathcal{F}'= g^{-1}\mathcal{F}g,\ \ \
	\mathcal{G}'= g^{-1}\vartriangleright \mathcal{G} + \mathcal{F}' \wedge^{\vartriangleright}\phi.
\end{align}

In the above expression, we have  mostly followed the notations and conventions of \cite{SDH5}. We denote the vector space of $\mathcal{g}$-valued differential $k$-forms on $M$ over $C^{\infty}(M)$ by $\Omega^k(M, \mathcal{g})$. The convention also applies to $\mathcal{h}$.  Here, we consider matrix Lie algebras, and define the exterior differential and wedge product of Lie algebra valued differential forms to be
\begin{align}
	dA := \sum\limits_{a} d A^a X_a,\ \ 
	A_1 \wedge A_2 := \sum \limits_{ab}A^a_1 \wedge A^b_2 X_a X_b,\ \ 
	A_1 \wedge^{[, ]}A_2 := \sum \limits_{ab}A^a_1 \wedge A^b_2 [X_a, X_b],\\
		\alpha(B):= \sum \limits_{b}B^b \alpha(Y_b),\ \ \ 
	A \wedge^{\vartriangleright}B:= \sum \limits_{ab} A^a \wedge B^b X_a \vartriangleright Y_b,
\end{align}
for $A_1 = \sum\limits_{a} A^a_1X_a \in \Omega^{k_1}(M, \mathcal{g})$, $A_2 = \sum\limits_{a} A^a_2X_a \in \Omega^{k_2}(M, \mathcal{g})$. 

Imitating the extended invariant form \eqref{ei}, there exists a higher invariant form 
\begin{align}\label{p}
	\mathcal{P}_{2n+3}=\langle \mathcal{F}^n, \mathcal{G}\rangle_{\mathcal{g}\mathcal{h}},
\end{align}
where $\langle \cdots \rangle_{\mathcal{g}\mathcal{h}}$ stands for a generalized multilinear symmetric invariant polynomial
for the differential crossed modules
$(\mathcal{h}, \mathcal{g}; \alpha, \vartriangleright)$, whose definition is given by \ref{H}.
It is easy to check that this form is gauge invariant under the higher gauge transformation \eqref{2gt11} and \eqref{2gt22}. Besides, it is closed, i.e., $d \langle \mathcal{F}^n, \mathcal{G} \rangle_{\mathcal{g}\mathcal{h}}=0$, by direct computation of the derivative. According to the Poincar\'{e} lemma, this implies that $	\mathcal{P}_{2n+3}$ can be locally written as an exterior differential of a certain $(2n+2)$-form, which can be given by the higher Chern-Weil theorem.

\textbf{Higher Chern-Weil theorem:}
Let $(A_0, B_0)$ and $(A_1, B_1)$ be two 2-connections, and the corresponding curvature forms are given by
\begin{align}
	\mathcal{F}_i= dA_i+ \dfrac{1}{2}A_i \wedge^{[, ]} A_i - \alpha(B_i), \ \ \ 
	\mathcal{G}_i= dB_i + A_i \wedge^{\vartriangleright}B_i,
\end{align}
for $i=0,1$.
Define the interpolations between the two connections 
\begin{align}
	A_t= A_0 + t \theta, \ \ \theta= A_1-A_0, \label{d1}\\
	B_t = B_0 + t \Phi,\ \ \Phi= B_1 - B_0,\label{d2}
\end{align}
for $0 \leq t \leq 1$ and their curvatures are given by
\begin{align}\label{c1}
	\mathcal{F}_t = dA_t + \dfrac{1}{2}A_t \wedge^{[, ]} A_t - \alpha(B_t),\ \ \ 
	\mathcal{G}_t= dB_t + A_t \wedge^{\vartriangleright}B_t.
\end{align}
Then, have
\begin{align}\label{ch1}
	\mathcal{P}^{(1)}_{2n+3}-\mathcal{P}^{(0)}_{2n+3}=\langle \mathcal{F}^n_1, \mathcal{G}_1 \rangle_{\mathcal{g}\mathcal{h}} - \langle \mathcal{F}^n_0, \mathcal{G}_0 \rangle_{\mathcal{g}\mathcal{h}}= d \mathcal{Q}^{2n+2}(A_0, B_0; A_1, B_1),
\end{align}
where
\begin{align}\label{2AST}
	\mathcal{Q}^{2n+2}(A_0, B_0; A_1, B_1)= \int_{0}^{1}dt\Big\{n\langle \theta \wedge \mathcal{F}^{n-1}_t, \mathcal{G}_t\rangle_{\mathcal{g}\mathcal{h}} + \langle \mathcal{F}^n_t, \Phi\rangle_{\mathcal{g}\mathcal{h}}\Big\}
\end{align}
 is called a 2-Antoniadis-Savvidy transgression form, which is a higher analog of the Antoniadis-Savvidy transgression form \eqref{AST}.

Setting $A_0=B_0=0$ and $A_1=A$, $B_1=B$ in \eqref{2AST}, one can get the 2-Chern-Simons-Antoniadis-Savvidy (2ChSAS) form
\begin{align}\label{QQ}
	\mathcal{Q}^{2n+2}(0, 0; A, B)=\int_{0}^{1}dt \Big\{n \langle A \wedge \mathcal{F}^{n-1}_t, \mathcal{G}_t \rangle_{\mathcal{g}\mathcal{h}} + \langle \mathcal{F}^n_t, B \rangle_{\mathcal{g}\mathcal{h}}\Big\}= \mathcal{C}^{2n+2}_{2ChSAS},
\end{align}
which satisfies \eqref{ch1} becoming
\begin{align}
	\langle \mathcal{F}^n, \mathcal{G}\rangle_{\mathcal{g}\mathcal{h}} = d \mathcal{C}^{2n+2}_{2ChSAS}.
\end{align}
Particularly, for $n=1$, there is the same result in \cite{SDH4}, where the author computed $Q_{2CS}$ as the $4$-d CS form 
\begin{align}
	\mathcal{Q}^{4}(0, 0; A, B)=\int_{0}^{1}dt \Big\{\langle A, \mathcal{G}_t \rangle_{\mathcal{g}\mathcal{h}} + \langle \mathcal{F}_t, B \rangle_{\mathcal{g}\mathcal{h}}\Big\}
	= Q_{2CS}.
\end{align}

We conclude similarly that the generalized Chern-Weil theorem can be generalized to the higher partner, whose particular case gives the explicit expression of the 2ChSAS form.
In other wards, the 2ChSAS form is regard as a particular case of the 2-Antoniadis-Savvidy transgression form, which is a more general object similar to the generalized transgression form \eqref{AST}.
This conclusion makes a preparation of the consideration of a subspace separation method for the 2ChSAS theory in section \ref{hECHF}.

\section{The extended Cartan homotopy formula for $ \Pi= \langle \mathcal{F}^n_t, \mathcal{G}_t\rangle_{\mathcal{g}\mathcal{h}}$}\label{hECHF}
From what has already been given for (generalized) CS theory, we obtain that the (generalized) ECHF can reproduce the (generalized) Chern-Weil theorem and triangle equation \cite{FEP, FIPSPSS}. These results encourage us to develop further the higher ECHF for the 2ChSAS theory \cite{SDH5}.
And we will show that it can also reproduce the higher Chern-Weil theorem and give the corresponding higher triangle equation. The latter states that we can find a subspace separation method to allow for a deeper understanding of the 2ChSAS Lagrangian too.

Let us consider a family of 2-connections $\{ (A_i, B_i), i=0, \cdots, p+1\}$ on the principal 2-bundle $E$ over the $(2n+2)$-dimensional manifold $M$. Let $T_{p+1}$ be a $(p+1)$-dimensional oriented simplex 
smoothly parametrized by a set of variables $\{ t^i, i=0, \cdots, p+1\}$, which must satisfy the constraints 
\begin{align}
	\sum_{i=0}^{p+1}t^i=1,\ \ 	t^i \geq 0, \ \ \ i=0, \cdots, p+1.
\end{align}
The simplex and its boundary are denoted by
\begin{align}
	T_{p+1}&=(A_0, B_0; \cdots; A_{p+1}, B_{p+1}),\\
	\partial T_{p+1}&= \sum_{i=0}^{p+1}(-1)^iT^{(i)}_{p}(A_0, B_0; \cdots; \hat{A}_i, \hat{B}_i; \cdots; A_{p+1}, B_{p+1}),
\end{align}
where the symbol `` $\hat{ }$ " over $A_i, B_i$ indicates that $A_i, B_i$ is deleted from the sequence $(A_0, B_0; \cdots; A_{p+1}, B_{p+1})$. 
It follows immediately that the convex combination $(A_t, B_t)$ with
\begin{align}
	A_t= \sum_{i=0}^{p+1}t^iA_i,\ \  B_t= \sum_{i=0}^{p+1}t^i B_i,
\end{align}
transforms as a gauge 2-connection in the same way as every individual $(A_i, B_i)$ does.

Apart from the usual antiderivation $d$ with respect to $x$, there is another antiderivation $d_t$ with respect to the parameter $t$. Besides, there is also a homotopy derivation $l_t$ given by
\begin{align}
	l_t: \Omega^a(M)\times \Omega^b(T)\longrightarrow \Omega^{a-1}(M)\times \Omega^{b+1}(T),
\end{align}
i.e., decreasing the degree in $dx$ by one and increasing the degree in $dt$ by one. It satisfies  Leibniz's rule as well as $d$ and $d_t$. The three operators $d$, $d_t$ and $l_t$ define a graded algebra satisfying
\begin{align}\label{grad}
	d^2= d^2_t= 0, \ \  d d_t+ d_t d=0,\ \ l_td-dl_t=d_t,\ \ l_td_t-d_tl_t=0.
\end{align}

For $f(l_t)$ a polynomial in $l_t$, it can be verified the relationship
\begin{align}\label{rela}
	[f(l_t), d]=d_tf'(l_t)=f'(l_t)d_t,
\end{align}
by using \eqref{grad}. Taking $f(l_t)=e^{l_t}$, as given by its Taylor expansion, we have
\begin{align}\label{e}
	e^{l_t}d-de^{l_t}=d_t e^{l_t}=e^{l_t} d_t
\end{align}
from \eqref{rela}. These results go back to the work of Zumino et al. \cite{Zumino}.

Let $\Pi$ be a polynomial in the forms $\{A_t, B_t, \mathcal{F}_t, \mathcal{G}_t, d_tA_t, d_tB_t, d_t\mathcal{F}_t, d_t\mathcal{G}_t\}$, where $\mathcal{F}_t$, $\mathcal{G}_t$ are the corresponding curvature forms
\begin{align}
	\mathcal{F}_t=dA_t + A_t \wedge^{[, ]} A_t - \alpha(B_t), \ \ \mathcal{G}_t=dB_t + A_t \wedge^{\vartriangleright}B_t.
\end{align}
The action of $l_t$ is defined by
\begin{align}
	l_t A_t=l_t B_t=0, 
\end{align}
which derives the relations
\begin{align}\label{1}
	l_t \mathcal{F}_t=d_t A_t,\ \  \ 
	l_t \mathcal{G}_t= d_t B_t.
\end{align}
Then, let Eq.\eqref{e} act on the polynomial $\Pi$, having
\begin{align}
	e^{l_t} d_t\Pi=e^{l_t}d \Pi-de^{l_t}\Pi.
\end{align}
Expanding both sides of this equation, we obtain
\begin{align}\label{diff}
	\dfrac{l^p_t}{p!}d_t\Pi =\dfrac{l^{p+1}_t}{(p+1)!}d \Pi-d\dfrac{l^{p+1}_t}{(p+1)!}\Pi,
\end{align}
which is just the higher ECHF in differential form. 
Integrating this equation over $T_{p+q+1}$ in the space of parameters $\{t^i \}$, we obtain the integral form of the higher ECHF
\begin{align}\label{ECHF3}
	\int_{\partial T_{p+q+1}}\dfrac{l^p_t}{p!}\Pi = \int_{T_{p+q+1}}\dfrac{l^{p+1}_t}{(p+1)!}d\Pi +(-1)^{p+q}d \int_{T_{p+q+1}}\dfrac{l^{p+1}_t}{(p+1)!}\Pi,
\end{align}
where $q$ is the degree of $\Pi$ on $T_{p+q+1}$. If $\Pi$ is an $m$-form on $M$, then we have $p \leq m$. About the convention for this ``incomplete" integration, see Ref. \cite{Zumino} for more details.

Let us consider the polynomial 
\begin{align}
	\Pi= \langle \mathcal{F}^n_t, \mathcal{G}_t\rangle_{\mathcal{g}\mathcal{h}},
\end{align}
which has the three following properties,
\begin{itemize}
	\item $\Pi$ is $M$-closed, i.e., $d \Pi =0$;
	\item $\Pi$ is a $0$-form on $T$,  i.e., $q=0$;
	\item $\Pi$ is a $(2n+3)$-form on $M$, i.e., $m=2n+3$.
\end{itemize}
Thus, we have $p$ with valued in $0, \cdots, 2n+3$ and the higher ECHF \eqref{ECHF3} reduces in this case to
\begin{align}\label{echf}
	\int_{\partial T_{p+1}}\dfrac{l^p_t}{p!} \langle \mathcal{F}^n_t, \mathcal{G}_t\rangle_{\mathcal{g}\mathcal{h}} = (-1)^p d \int_{T_{p+1}}\dfrac{l^{p+1}_t}{p+1}\langle \mathcal{F}^n_t, \mathcal{G}_t\rangle_{\mathcal{g}\mathcal{h}},
\end{align}
which is also called a set of higher descent equations.

In the following subsection, we will study two cases $p=0$ and $p=1$ of the Eq. \eqref{echf}.
As what we have anticipated, the higher ECHF is the common origin of the higher Chern-Weil theorem and the triangle equation.
\subsection{$p=0$: Higher Chern-Weil theorem}\label{hECHF1}
In this subsection, we study the case $p=0$ of Eq. \eqref{echf} reading 
\begin{align}\label{cw}
	\int_{\partial T_1} \langle \mathcal{F}^n_t, \mathcal{G}_t\rangle_{\mathcal{g}\mathcal{h}} = d \int_{T_1} l_t \langle \mathcal{F}^n_t, \mathcal{G}_t\rangle_{\mathcal{g}\mathcal{h}},
\end{align} 
where $\mathcal{F}_t$ and $\mathcal{G}_t$ are the higher curvature forms for the 2-connection $(A_t, B_t)$ defined by
\begin{align}
	A_t&= A_0 + t\theta,\ \ \theta= A_1-A_0,\\
	B_t&= B_0 + t \Phi,\ \ \Phi=B_1-B_0.
\end{align}
The simplex is denoted by $T_1=(A_0, B_0; A_1, B_1)$, and its boundary is just
\begin{align}
	\partial T_1=(A_1, B_1)- (A_0, B_0).
\end{align}

Clearly, the integration of the left of \eqref{cw} is  given by
\begin{align}
	\int_{\partial T_1} \langle \mathcal{F}^n_t, \mathcal{G}_t\rangle_{\mathcal{g}\mathcal{h}} 
	=
	\langle \mathcal{F}^n_1, \mathcal{G}_1 \rangle_{\mathcal{g}\mathcal{h}} - \langle \mathcal{F}^n_0, \mathcal{G}_0 \rangle_{\mathcal{g}\mathcal{h}}.
\end{align}
On the other hand, we have 
\begin{align}
	l_t\mathcal{F}_t= \theta dt, \ \ l_t\mathcal{G}_t= \Phi dt,
\end{align}
by using \eqref{1}, and the symmetric nature of $\langle \mathcal{F}^n_t, \mathcal{G}\rangle_{\mathcal{g}\mathcal{h}}$ implies that
\begin{align}
	l_t \langle \mathcal{F}^n_t, \mathcal{G}_t\rangle_{\mathcal{g}\mathcal{h}}=n\langle	l_t\mathcal{F}_t\wedge \mathcal{F}^{n-1}_t, \mathcal{G}_t\rangle_{\mathcal{g}\mathcal{h}} + \langle \mathcal{F}^n_t, l_t\mathcal{G}_t \rangle_{\mathcal{g}\mathcal{h}}.
\end{align}
Thus, the Eq. \eqref{cw} finally becomes
\begin{align}
  \langle \mathcal{F}^n_1, \mathcal{G}_1 \rangle_{\mathcal{g}\mathcal{h}} - \langle \mathcal{F}^n_0, \mathcal{G}_0 \rangle_{\mathcal{g}\mathcal{h}}=d \mathcal{Q}^{2n+2}(A_0, B_0; A_1, B_1),
\end{align}
where $\mathcal{Q}^{2n+2}(A_0, B_0; A_1, B_1)$ is the 2-Antoniadis-Savvidy transgression form in \eqref{2AST}.

We would like to stress that the use of the higher ECHF has also allowed us to pinpoint the exact form of the higher transgression
\begin{align}\label{22}
	\mathcal{Q}^{2n+2}(A_0, B_0; A_1, B_1)=\int_{T_1} l_t \langle \mathcal{F}^n_t, \mathcal{G}_t\rangle_{\mathcal{g}\mathcal{h}}.
\end{align}
 This concludes that the higher Chern-Weil theorem derivated in \cite{SDH5} can be regarded as a corollary of the higher ECHF \eqref{echf}.

\subsection{$p=1$: Higher triangle equation}\label{hECHF2}
In this subsection, let us consider the second particular case $p=1$ of \eqref{echf} reading
\begin{align}\label{hte}
	\int_{\partial T_{2}} l_t \langle \mathcal{F}^n_t, \mathcal{G}_t\rangle_{\mathcal{g}\mathcal{h}} = -d \int_{T_{2}} \dfrac{l^2_t}{2} \langle \mathcal{F}^n_t, \mathcal{G}_t\rangle_{\mathcal{g}\mathcal{h}},
\end{align}
where $\mathcal{F}_t$ and $\mathcal{G}_t$ are the higher curvatures for the 2-connection $(A_t, B_t)$ defined by
\begin{align}
	A_t= A_0 + t^1\theta_{10} + t^2\theta_{20},\ \ \ 
	B_t= B_0 + t^1\Phi_{10} + t^2\Phi_{20},
\end{align}
with $\theta_{i0}=A_i-A_0$ and $\Phi_{i0}=B_i-B_0$ for $i=1,2$.
In this situation, the simplex becomes $T_2=(A_0, B_0; A_1, B_1; A_2, B_2)$, and the boundary is just
\begin{align}\label{boundary}
	\partial T_2= T^{(0)}_1- T^{(1)}_1 + T^{(2)}_1,
\end{align}
with $T^{(0)}_1= (A_1, B_1; A_2, B_2)$, $T^{(1)}_1=(A_0, B_0; A_2, B_2)$ and $T^{(2)}_1=(A_0, B_0; A_1, B_1)$.

In addition, we have
\begin{align}\label{y}
	l_t\mathcal{F}_t= \theta_{10} dt^1+ \theta_{20} dt^2, \ \ \ l_t\mathcal{G}_t= \Phi_{10} dt^1+ \Phi_{20} dt^2
\end{align}
by using \eqref{1}. According to the definition of
the boundary \eqref{boundary}, it follows that the left side of \eqref{hte} is decomposed as
\begin{align}
	 \int_{\partial T_{2}} l_t \langle \mathcal{F}^n_t, \mathcal{G}_t\rangle_{\mathcal{g}\mathcal{h}} =\mathcal{Q}^{2n+2}(A_1,B_1; A_2, B_2)-\mathcal{Q}^{2n+2}(A_0,B_0; A_2, B_2) + \mathcal{Q}^{2n+2}(A_0, B_0; A_1, B_1),
\end{align}
where each of the right terms in this equation is a 2-Antoniadis-Savvidy transgression form.

On the other hand, the Leibniz's rule for $l_t$ and \eqref{y} imply that
\begin{align}
-\int_{T_{2}}\dfrac{l^2_t}{2}\langle \mathcal{F}^n_t, \mathcal{G}_t\rangle_{\mathcal{g}\mathcal{h}} &=- \frac{n(n-1)}{2}\int_{T_{2}}\langle (l_t \mathcal{F}_t)^2 \wedge \mathcal{F}^{n-2}_t, \mathcal{G}_t \rangle_{\mathcal{g}\mathcal{h}} - n \int_{T_{2}}\langle l_t \mathcal{F}_t \wedge \mathcal{F}^{n-1}_t, l_t \mathcal{G}_t \rangle_{\mathcal{g}\mathcal{h}}\nonumber\\
	&= \int_{0}^{1} dt^1 \int_{0}^{1-t^1}dt^2   \Big\{n(n-1) \langle (A_1 -A_0)\wedge(A_2
	-A_0)\wedge \mathcal{F}^{n-2}_t, \mathcal{G}_t \rangle_{\mathcal{g}\mathcal{h}}  \nonumber\\
	& + n\langle (A_1-A_0) \wedge \mathcal{F}^{n-1}_t, (B_2-B_0) \rangle_{\mathcal{g}\mathcal{h}}-n\langle (A_2-A_0) \wedge \mathcal{F}^{n-1}_t, B_1-B_0 \rangle_{\mathcal{g}\mathcal{h}}\Big\}\nonumber\\
	&=	\mathcal{Q}^{2n+1}(A_0,B_0; A_1, B_1; A_2, B_2).
\end{align}

Putting everything together, we get the higher triangle equation
\begin{align}\label{2}
	\mathcal{Q}^{2n+2}(A_0,B_0; A_2, B_2)
	=& \mathcal{Q}^{2n+2}(A_1,B_1; A_2, B_2) + \mathcal{Q}^{2n+2}(A_0, B_0; A_1, B_1) \nonumber\\
	&- d \mathcal{Q}^{2n+1}(A_0,B_0; A_1, B_1; A_2, B_2).
\end{align}
Similar to the ordinary triangle equation in \cite{FIERPS}
and the generalized  triangle equation  in \cite{FIPSPSS}, we can also pinpoint the exact form of the boundary contribution $\mathcal{Q}^{2n+1}(A_0,B_0; A_1, B_1; A_2, B_2)$ by using the higher ECHF.
Because of the fact that the 2ChSAS form is regard as a particular case of the 2-Antoniadis-Savvidy transgression form, we obtain an expression that relates the 2-Antoniadis-Savvidy transgression form to two 2ChSAS forms and a total derivative
\begin{align}
		\mathcal{Q}^{2n+2}(A_0,B_0; A_1, B_1)
	=\mathcal{C}^{2n+2}_{2ChSAS}(A_1, B_1)-\mathcal{C}^{2n+2}_{2ChSAS}(A_0, B_0)+d \mathcal{Q}^{2n+1}(A_0,B_0; A_1, B_1; 0, 0; )
\end{align}
by choosing $A_2=B_2=0$ in \eqref{2}. This conclusion can also be derived through Cartan homotopy formula for the 2ChSAS form in the following section.

Thus, it is possible that there is also a separation method based on the higher triangle equation \eqref{2}, which allows one to separate the 2ChSAS action in bulk and boundary contributions, and split the Lagrangian in appropriate reflection of the subspace structure of the gauge algebra, systematically. 
This method shall follow by the same steps as in ordinary CS theory \cite{FIERPS}, the only difference being in the analysis of the gauge algebra.
But we will not develop this point here. 

\section{The Cartan homotopy formula for $	\Pi=\mathcal{C}^{2n+2}_{2ChSAS}(A_t, B_t)$}\label{hCHF}
In Ref. \cite{PRRJ}, one study the relation between the transgression forms and the CS forms by using the Cartan homotopy formula.  
In this section, we adopt the same technique to investigate the 2ChSAS theory. 

It is known that the ECHF includes as a special case the ordinary Cartan homotopy formula.
The same reasoning applies to the higher ECHF \eqref{diff}, thus we consider the particular case $p=0$
   \begin{align}
	d_t\Pi =l_td \Pi-d l_t \Pi,
	\end{align}
which we call a higher Cartan homotopy formula in differential, and its integral form is given by
\begin{align} \label{chf}
		\int_{\partial T_{q+1}} \Pi = \int_{T_{q+1}}l_t d\Pi +(-1)^{q}d \int_{T_{q+1}}l_t\Pi.
\end{align}
The higher Chern-Simons theorem will be reproduced if we still take $\Pi=\langle \mathcal{F}^n_t, \mathcal{G}_t \rangle_{\mathcal{g}\mathcal{h}}$ in the above equation. This brings us back to the discussion of the subsection \ref{hECHF1}.

 Our task now is to consider 
\begin{align}
	\Pi=\mathcal{C}^{2n+2}_{2ChSAS}(A_t, B_t)
	=\int_{0}^{1}ds \Big\{n \langle A_t \wedge \mathcal{F}^{n-1}_{st}, \mathcal{G}_{st} \rangle_{\mathcal{g}\mathcal{h}} + \langle \mathcal{F}^n_{st}, B_t \rangle_{\mathcal{g}\mathcal{h}}\Big\},
\end{align}
where
\begin{align}
	A_t=A_0+t \theta,\ \ \ \mathcal{F}_{st}=s\mathcal{F}_t+(s^2-s)A_t\wedge A_t,\\
	B_t=B_0+t \Phi,\ \ \ \mathcal{G}_{st}=s\mathcal{G}_t+(s^2-s)A_t\wedge^{\vartriangleright} B_t.
\end{align}
It is clear that $d \Pi = \langle \mathcal{F}^n_t, \mathcal{G}_t \rangle_{\mathcal{g}\mathcal{h}}$ and $\Pi$ is a $0$-form on the simplex $T_1$,  i.e., $q=0$.


In this case, the equation \eqref{chf} reads 
\begin{align}\label{12}
	\int_{\partial T_1}\mathcal{C}^{2n+2}_{2ChSAS}(A_t, B_t)=\int_{T_1}l_t d\mathcal{C}^{2n+2}_{2ChSAS}(A_t, B_t)+d \int_{T_1}l_t \mathcal{C}^{2n+2}_{2ChSAS}(A_t, B_t)
\end{align}
Evidently, one can observe that
\begin{align}
	\int_{\partial T_1}\mathcal{C}^{2n+2}_{2ChSAS}(A_t, B_t)=\mathcal{C}^{2n+2}_{2ChSAS}(A_1, B_1)-\mathcal{C}^{2n+2}_{2ChSAS}(A_0, B_0),
\end{align}
and 
\begin{align}
	\int_{T_1}l_t d\mathcal{C}^{2n+2}_{2ChSAS}(A_t, B_t)=\int_{T_1}l_t \langle \mathcal{F}^n_t, \mathcal{G}_t\rangle_{\mathcal{g}\mathcal{h}} =\mathcal{Q}^{2n+2}(A_0, B_0; A_1, B_1),
\end{align}
by using \eqref{22}.
On the other hand, we have
\begin{align}
	B_{2n+1}&=\int_{T_1}l_t \mathcal{C}^{2n+2}_{2ChSAS}(A_t, B_t)\nonumber\\
	&=\int_{T_1}l_t\int_{0}^{1}ds \Big\{n \langle A_t \wedge \mathcal{F}^{n-1}_{st}, \mathcal{G}_{st} \rangle_{\mathcal{g}\mathcal{h}} + \langle \mathcal{F}^n_{st}, B_t \rangle_{\mathcal{g}\mathcal{h}}\Big\}\nonumber\\
	&=-\int_{0}^{1}dt \int_{0}^{1}ds ns \Big\{ (n-1)\langle A_t \wedge \theta \wedge \mathcal{F}^{n-2}_{st}, \mathcal{G}_{st}\rangle_{\mathcal{g}\mathcal{h}} + \langle A_t \wedge \mathcal{F}^{n-1}_{st}, \Phi \rangle_{\mathcal{g}\mathcal{h}} - \langle \theta \wedge \mathcal{F}^{n-1}_{st}, B_t \rangle_{\mathcal{g}\mathcal{h}} \Big\}.
\end{align}
 Therefore, we infer that a 2-Antoniadis-Savvidy transgression form can be written as the difference of two 2ChSAS forms minus an exact form, i.e.,
 \begin{align}\label{cc}
 \mathcal{Q}^{2n+2}(A_0, B_0; A_1, B_1)=	\mathcal{C}^{2n+2}_{2ChSAS}(A_1, B_1)-\mathcal{C}^{2n+2}_{2ChSAS}(A_0, B_0)-dB_{2n+1}
 \end{align}
from \eqref{12}.

There is a direct consequence that the exterior derivative of the 2-Antoniadis-Savvidy transgression form is the difference of two higher invariant forms \eqref{p}.
In Ref. \cite{SDH5}, we know that $\mathcal{Q}^{2n+2}(A_0, B_0; A_1, B_1)$ is gauge invariant under the general 2-gauge transformation \eqref{2gt11} and \eqref{2gt22}. Therefore, the role of the surface term $B_{2n+1}$ is to cancel the variation of the bulk terms $\mathcal{C}^{2n+2}_{2ChSAS}$, which change by a closed form under the 2-gauge transformation.

\section{Concluding remarks}\label{Concluding}
In this Letter, we constructed the higher ECHF based on the 2-gauge theory, and gave its differential and integral forms respectively. Then, we studied a particular case $\Pi=\mathcal{P}_{2n+3}$, and obtained a set of higher descent equations. 
Ulteriorly, we showed that the higher ECHF can recover the higher Chern-Weil theorem and yield the higher triangle equation. 
Finally, we developed the higher Cartan homotopy formula, which is regarded as a special case of the higher ECHF. Letting $\Pi=\mathcal{C}^{2n+2}_{2ChSAS}(A_t, B_t)$, we proved that a 2-Antoniadis-Savvidy transgression form can be written as the difference of two 2ChSAS forms minus an exact form.
Based on these findings, future work will focus on two concerns:

On the one hand, higher transgression gauge field theory was established in \cite{SDH5}. Meanwhile, \cite{FIERPS} detailed the subspace separation method for standard transgression theory. A natural research extension would be to generalize this separation method. Specifically, we propose applying it to higher transgression field theory. This could open new applications in mathematical physics.

On the other hand, the ECHF establishes the theoretical basis for descent equations, which play a fundamental role in gauge and topological quantum field theories \cite{Zumino}. For example, these equations establish systematic relations between higher-dimensional topological invariants (e.g., $Tr(F^k)$) and lower-dimensional differential forms (such as Chern-Simons forms and anomaly polynomials), thereby localizing global topological information. Moreover, the descent equations rigorously explain chiral anomalies and addresses key issues in quantum gauge theory, including BV quantization and WZW model construction \cite{ACCR, MG07}. This study found that higher ECHF yields corresponding higher descent equations, and this extended formula has been successfully applied to the strict higher Chern-Simons theory. This raises two questions:
	\begin{enumerate}
		\item Can the derivation of the WZW model as the boundary theory of the strict higher Chern-Simons theory and its quantization be systematically derived from the higher ECHF?
		\item Can this method handle the 4-dimensional semistrict Chern-Simons theory \cite{R.Z4, R.Z5}, including unresolved challenges like gauge-fixing covariance with ``ghost-for-ghost" and quantum anomalies?
	\end{enumerate}
The quantization of higher Chern-Simons theory, a subject of growing interest in contemporary theoretical physics, has attracted considerable attention due to its fundamental significance and potential applications in diverse fields. Notably, the higher ECHF may provide a crucial breakthrough in addressing this challenging problem.

\section*{Acknowledgment}
This work is supported by the National Natural Science Foundation of China (Nos.11871350).

\appendix

\section{Lie crossed modules and differential crossed modules}\label{crossed module}
In this appendix, we give the basic definitions and relations in order to define our terminologies and notations and for reference throughout in the text.
See Ref. \cite{Martins:2010ry} for more details.

\textbf{Lie crossed module:}
	A crossed module $(H, G; \bar{\alpha}, \bar{\vartriangleright})$ is given by a group morphism $\bar{\alpha}: H \longrightarrow G$ together with a left action $\bar{\vartriangleright}$ of $G$ on $H$ by automorphisms, such that:
	\begin{itemize}
		\item $\bar{\alpha}(g \bar{\vartriangleright} h)=g \bar{\alpha}(h)g^{-1}$, for each $g \in G$ and $h \in H$;
		\item $\bar{\alpha}(h)\bar{\vartriangleright} h'=h h' h^{-1}$, for each $h, h'\in H$.
	\end{itemize}
If $G$ and $H$ are both Lie groups, and $\alpha$ is a smooth morphism, and the left action of $G$ on $H$ is smooth, then $(H, G; \bar{\alpha}, \bar{\vartriangleright})$ will be called a Lie crossed module.

Given a Lie crossed module $(H, G; \bar{\alpha}, \bar{\vartriangleright})$, then the induced Lie algebra is called a differential crossed module in the sense of the following definition (see Refs. \cite{BS1, BS2, B1}).

\textbf{Differential crossed module:} A differential crossed module $(\mathcal{h}, \mathcal{g}; \alpha, \vartriangleright)$ is given by a Lie algebra morphism $\alpha: \mathcal{h}\longrightarrow \mathcal{g}$ together with a left action of $\mathcal{g}$ on the underlying vector space of $\mathcal{h}$, such that:
\begin{itemize}
	\item For any $X \in \mathcal{g}$ the  map $Y \in \mathcal{h}\longrightarrow X \vartriangleright Y \in \mathcal{h}$ is a derivation of $\mathcal{h}$, which can be written as 
	\begin{align}
		X \vartriangleright[Y, Y']=[X\vartriangleright Y, Y']+[Y, X \vartriangleright Y'],\ \ \forall X\in \mathcal{g}, \forall Y, Y' \in \mathcal{h}.
	\end{align} 
\item The map $\mathcal{g}\longrightarrow Der(\mathcal{h})$ from $\mathcal{g}$  into the derivation algebra of $\mathcal{h}$ induced by the action of $\mathcal{g}$ on $\mathcal{h}$ is a Lie algebra morphism, which can be written as 
\begin{align}
	[X, X']\vartriangleright Y=X \vartriangleright(X' \vartriangleright Y)- X' \vartriangleright(X \vartriangleright Y), \ \ \forall X, X' \in \mathcal{g}, \forall Y \in \mathcal{h}.
\end{align}
\item 
\begin{align}
	\alpha(X \vartriangleright Y)=[X, \alpha(Y)], \ \ \forall X \in \mathcal{g}, \forall Y \in \mathcal{h}.
\end{align}
\item 
\begin{align}
	\alpha(Y)\vartriangleright Y' = [Y, Y'],\ \ \forall Y, Y' \in \mathcal{h}.
\end{align}
\end{itemize}

There are many examples of Lie 2-groups or Lie 2-algebras for applications in \cite{Baez.2010, Martins:2010ry,TM}. For example, the BFCG pure gravity is based on the Poincar$\acute{e}$ 2-group $(\mathbb{R}^4, SO(3,1); \bar{\alpha}, \bar{\vartriangleright})$. The map $\bar{\alpha}$ is trivial, and $ \bar{\vartriangleright}$ is the representation of $SO(3,1)$ on $\mathbb{R}^4$. The Poincar$\acute{e}$ 2-algebra is $(\mathbb{R}^4, \mathcal{so}(3,1); \alpha, \vartriangleright)$, with the generators $P_a$ and $M_{ab}$, respectively. The action $\vartriangleright$ is defined by
	\begin{align}
		M_{ab}\vartriangleright M_{cd}&=[M_{ab}, M_{cd}]=\eta_{ad}M_{bc}+\eta_{bc}M_{ad}-\eta_{ac}M_{bd}-\eta_{bd}M_{ac},\\
		M_{ab}\vartriangleright P_c&=\eta_{bc}P_a - \eta_{ac}P_b,
	\end{align}
where the indices $a, b, ...$ are the Lorentz indices, running from $0, ..., 3$, and $\eta_{ab}=diag(-1, +1, +1, +1)$ is the Minkowski metric.

\section{Multilinear symmetric invariant polynomial on $(\mathcal{h}, \mathcal{g}; \alpha, \vartriangleright)$}\label{H}

Based on the 2-gauge theory, one can define a generalized multilinear symmetric invariant polynomial
for the differential crossed module $(\mathcal{h}, \mathcal{g}; \alpha, \vartriangleright)$ in \cite{SDH5}
\begin{align}\label{mp}
	\langle \cdots, \cdot \rangle_{\mathcal{g}\mathcal{h}}: \mathcal{g}^n \times \mathcal{h}\longrightarrow \mathbb{R},
\end{align}
satisfying
\begin{align}
	&\langle X_1 \cdots  X_i \cdots X_n, X \vartriangleright Y\rangle_{\mathcal{g}\mathcal{h}}=-\sum_{i=1}^{n} \langle X_1 \cdots  [X, X_i] \cdots X_n, Y\rangle_{\mathcal{g}\mathcal{h}},\label{sy2}\\
	&\langle X_1 \cdots \alpha(Y_i) \cdots X_n, Y \rangle_{\mathcal{g}\mathcal{h}} = \langle X_1 \cdots \alpha(Y) \cdots X_n, Y_i\rangle_{\mathcal{g}\mathcal{h}}.\label{sy3}
\end{align}
The symmetry implies that 
\begin{align}\label{sy4}
	\langle X_1\cdots X_i \cdots X_j \cdots X_n, Y \rangle_{\mathcal{g}\mathcal{h}}=\langle X_1\cdots X_j \cdots X_i \cdots X_n, Y \rangle_{\mathcal{g}\mathcal{h}},
\end{align}
and the invariance states clearly that \begin{align}\label{inv}
	\langle g X_1g^{-1} \cdots g X_n g^{-1},  g\vartriangleright Y\rangle_{\mathcal{g}\mathcal{h}}=\langle X_1\cdots X_n, Y \rangle_{\mathcal{g}\mathcal{h}},
\end{align}
for each $g \in G$, which can be given by taking $g$ as an infinitesimal transformation and using the identity \eqref{sy2}. In the case of $n=1$, \eqref{mp} becomes a bilinear form $\langle \cdot, \cdot \rangle_{\mathcal{g}\mathcal{h}} : \mathcal{g}\times \mathcal{h}\longrightarrow \mathbb{R}$ in \cite{SDH4, R.Z5}.

Besides, the equation \eqref{sy2} essentially boils down to
\begin{align}\label{Dd}
	\langle D(A_1\wedge \cdots \wedge A_n, \hat{B})\rangle_{\mathcal{g}\mathcal{h}}=d \langle A_1\wedge \cdots \wedge A_n, \hat{B}\rangle_{\mathcal{g}\mathcal{h}},
\end{align}
where $\{ A_i, i=1, \cdots, n\}$ is a set of $\mathcal{g}$-valued differential forms and $\hat{B}$ is an $\mathcal{h}$-valued differential form.
Moreover, the symmetry requirement for $\langle \cdots, \cdot\rangle_{\mathcal{g}\mathcal{h}}$ implies that, for any $p$-form $P$ and $q$-form $Q$ valued in $\mathcal{g}$, we have
\begin{align}
	\langle \cdots P \cdots Q \cdots, \cdot\rangle_{\mathcal{g}\mathcal{h}}=(-1)^{pq}\langle \cdots Q \cdots P \cdots, \cdot\rangle_{\mathcal{g}\mathcal{h}}.
\end{align}
	
\end{document}